\documentclass[twocolumn,secnumarabic,amssymb, nobibnotes,prd,superscriptaddress,floatfix]{revtex4}

\usepackage{graphicx}
\usepackage{dcolumn}
\usepackage{bm}
\usepackage{color}
\usepackage{amsmath}
\usepackage{caption}
\usepackage{subcaption}
\usepackage{ragged2e}
\usepackage{soul}
\usepackage{multirow}
\usepackage[normalem]{ulem}

\begin{document}

\preprint{APS/123-QED}

\title{Multi-messenger Study of Galactic Diffuse Emission with LHAASO and IceCube Observations}

\author{Chengyu Shao} 
\affiliation{School of Physics and Astronomy, Sun Yat-Sen University, No.2 Daxue Rd, 519082, Zhuhai China}

\author{Sujie Lin} \email{linsj6@mail.sysu.edu.cn}
\affiliation{School of Physics and Astronomy, Sun Yat-Sen University, No.2 Daxue Rd, 519082, Zhuhai China}

\author{Lili Yang} \email{yanglli5@mail.sysu.edu.cn}
\affiliation{School of Physics and Astronomy, Sun Yat-Sen University, No.2 Daxue Rd, 519082, Zhuhai China}
\affiliation{Centre for Astro-Particle Physics, University of Johannesburg, PO Box 524, Auckland Park 2006, South Africa}

\date{\today}

\begin{abstract}
With the breakthrough in PeV gamma-ray astronomy brought by the LHAASO experiment, the high-energy sky is getting richer than before. Lately, LHAASO Collaboration reported the observation of a gamma-ray diffuse emission with energy up to the PeV level from both the inner and outer Galactic plane. In these spectra, there is one bump that is hard to explain by the conventional cosmic-ray transport scenarios. Therefore, we introduce two extra components corresponding to unresolved sources with exponential-cutoff-power-law (ECPL) spectral shape, one with an index of 2.4, and 20 TeV cutoff energy, and another with index of 2.3 and 2 PeV cutoff energy. With our constructed model, we simulate the Galactic diffuse neutrino flux and find our results are in full agreement with the latest IceCube Galactic plane search. We estimate the Galactic neutrino contributes of $\sim 9\%$ of astrophysical neutrinos at 20 TeV. In the high-energy regime, as expected most of the neutrinos observed by IceCube should be from extragalactic environments.

\end{abstract}

\maketitle


\section{Introduction}\label{intro} 
The origin of cosmic rays (CRs) is one of the key questions in astrophysics. The CR energy spectrum shows the knee and ankle features. It is generally believed that CRs with energies below the spectral knee at $\sim 10^{15}$ eV, mainly come from our Galaxy, so-called Galactic cosmic rays (GCRs). While those with energies above the spectral ankle at $\sim 10^{18}$ eV are mostly from extra-galactic energetic sources. Most CR particles may lose their directional information due to the deflection and interaction with extra-galactic and Galactic magnetic fields and medium during their propagation. This additional uncertainty means that the origin of CRs near the knee remains mysterious.

To resolve the puzzles, alternative methods have been adopted. Collisions between energetic CRs and ambient and interstellar medium generate neutral ($\pi ^{0}$) and charged pions ($\pi ^{\pm}$), which decay to gamma rays and neutrinos. These secondary products detected on Earth will encode details of both the CR and target populations. The accurate interpretation of such measurements can provide direct information on the propagation and sources of CRs.

In the last few decades, progress has been made in detecting high-energy gamma-ray and neutrino emissions. The continuum diffuse gamma-ray emission has been well measured by Fermi Large Area Telescope (LAT) up to a few hundred GeV~\cite{ackermann2012fermi, ackermann2015spectrum}. Later on, in the TeV energy regime, Milagro~\cite{Abdo:2008if}, ARGO-YBJ~\cite{ARGO-YBJ:2015cpa}, H.E.S.S.~\cite{aharonian2008energy, HESS:2017tce, abdalla2018hess} and HAWC~\cite{aab2019probing, albert20203hwc} have been contributing data in the Galactic plane. These measurements have only recently reached to PeV range thanks to the Tibet AS$\gamma$ and LHAASO~\cite{TibetASgamma:2021tpz, LHAASO:2021gok, LHAASO:2023gne}. This discovery suggests the existence of PeVatrons~\cite{Sudoh:2022sdk}, which are sources capable of accelerating particles up to PeV energies. It is absolutely a big step towards understanding cosmic-ray physics by exploring the knee region of the CR spectrum. 

On the other hand, since the first detection of astrophysical neutrinos in 2012, IceCube has been accumulating neutrino data for more than 10 years~\cite{IceCube:2017trr, IceCube:2019lzm, ANTARES:2018nyb}. With the development of machine learning techniques and more statistics, the neutrino emission from the Galactic plane has been identified lately~\cite{IceCube:diffuse}. 

All these achievements can provide a hint to the injection, distribution and propagation of CRs in our Galaxy. However, the analysis of the Galactic diffuse emission (GDE) can be seriously contaminated by unresolved Galactic point sources which may have a distribution similar to the interstellar gas. This brings a challenge to recognize the accelerator of CRs. Previously, a few groups have performed the studies on the diffuse emission from TeV to PeV, where they discussed the possibility of the Galactic diffuse gamma-ray and neutrino emission coming from cosmic-ray interaction, known sources, and unresolved sources~\cite{Luque:2022buq, Egberts:2023dmi, DeLaTorreLuque:2023wwg, Schwefer:2022zly, Gabici:2019jvz}.

In this work, based on the current cosmic-ray and Fermi-LAT data, and the most recent LHAASO and IceCube Galactic plane observation, we apply the popular GALPROP code~\cite{Strong:1998pw} to model CR transport and generate simulated spectra and maps of the diffuse gamma-ray and neutrino emissions. Specifically, we adopt a Diffusion plus Reacceleration (DR) model, and employ DR-high and DR-low models to take into account the uncertainties of the measurements obtained from the ground-based air-shower experiments, IceTop and KASCADE respectively. However, we find a tension between the predicted gamma-ray flux with our constructed models and the observations. To illustrate the characteristic of the LHAASO Galactic plane spectrum and explain the excess, we invent two populations of Galactic sources (EXTRA1 and EXTRA2) with exponential-cutoff-power-law (ECPL) spectra shape. In the energy up to $10^5~\mathrm{GeV}$, the spectrum of EXTRA1 has an index of 2.40 and $\sim$ 20 TeV cutoff energy. In the higher energy end up to $10^6~\mathrm{GeV}$, another component EXTRA2 with an index of 2.3 and a 2 PeV cutoff is introduced. This can be naturally explained by the two types of unresolved sources in our Galaxy with different maximum cosmic-ray energy. 

Based on the constructed models, we also estimate the diffuse Galactic neutrino flux that is consistent with the latest IceCube Galactic plane search. We found the Galactic neutrinos can contribute around 9 percent to the all-sky neutrino events at 20 TeV. At PeV energy, most of the neutrinos are coming from outside of our Galaxy. However, due to a few uncertain factors like the mechanisms, numbers, and distribution of these unresolved sources in our Galaxy and limited observation capability, there is still some space allowed for modeling. Therefore, to further reveal the puzzles, the next-generation Imaging Air Cherenkov Telescopes (IACTs) and neutrino detectors with a larger effective area and better angular and energy resolution, which can dramatically provide a precise location and morphology of sources, are in high demand. 

The paper is organized as follows. Section~\ref{data} provides the description of the multi-messenger data, including cosmic-ray, gamma-ray, and neutrino observations used in this work. In Section~\ref{sec:models}, we present the injection and propagation models of cosmic rays, together with the addition of extra source components for fitting the gamma-ray data. Based on the constructed models, we show the calculated Galactic diffuse gamma-ray and neutrino emission in Section~\ref{resutls}. In Section~\ref{discussion}, we give a discussion about the obtained results and the origins of the two extra components. In Section~\ref{summary}, we give a summary and future outlook for the multi-messenger observations.

\section{Multi-messenger observation}\label{data} 
Thanks to the development of both satellites and ground-based observatories, diffuse high-energy neutrinos with energies between 10 TeV to PeV~\cite{IceCube:2013low}, ultra-high-energy cosmic rays (UHECRs, $> 10^{18}$ eV) ~\cite{PierreAuger:2015fol}, and high-energy gamma rays from MeV to PeV have been measured, or upper limits have been provided~\cite{Fermi-LAT:2009ihh, aab2019probing, LHAASO:2023gne}.
As there is a natural connection between these three messengers, neutrinos and gamma rays are produced during the CR propagation and can directly point back to the origin of CRs. Their joint detection and analysis should be a very efficient way to explore the Universe. ~\cite{IceCube:2018dnn, LIGOScientific:2017zic}.
Moreover, the energy budgets of UHECRs, PeV neutrinos, and isotropic sub-TeV gamma rays are comparable~\cite{Fang:2017zjf}, which supports the unification of high-energy cosmic particles.

Before exploring, understanding, and identifying the mechanisms and physical processes of the astrophysical sources of CRs, the diffuse backgrounds originating from our Galaxy should be seriously studied. One accurate diffuse template can provide great help in analyzing the upcoming data. For this purpose, we attempt to constrain the diffuse emission with current observation. The measurements used in this work are presented below.

\subsection{High-energy cosmic rays}
The high-energy CR particles are accelerated by energetic astrophysical sources like supernova remnants (SNRs) and propagate inside the Galactic magnetic field around the Galactic disk after escaping. Although only the CR fluxes around the sun could be measured, their distribution throughout the Galaxy can be predicted by the propagation model. Generally, the propagation model is constrained by the secondary-to-primary flux ratio observation, such as B/C~\cite{dampecollaborationDetectionSpectralHardenings2022} and $\mathrm{^{10}Be/^9Be}$~\cite{AMS:2016brs}. More details regarding the propagation model can be found in Section~\ref{sec:models}.

Their fluxes around the earth have been directly measured by space-born experiments like AMS-02~\cite{AMS:2015tnn, AMS:2016brs, AMS:2017seo, AMS:2019iwo, AMS:2019rhg} and DArk  Matter  Particle  Explorer (DAMPE)~\cite{Alemanno:2021gpb,an_measurement_2019,ambrosi_direct_2017,dampecollaborationDetectionSpectralHardenings2022}, and also indirectly measured by the ground-based experiments like IceTop~\cite{IceCube:2019hmk} and KASCADE~\cite{KASCADE:2005ynk}.

\begin{figure}[!htp]
\centering
\begin{subfigure}[b]{0.45\textwidth}
\includegraphics[width=\textwidth]{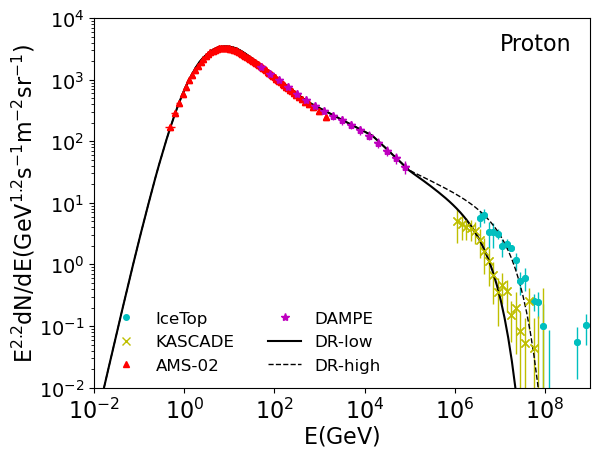}
\caption{proton}
\label{fig:elements:H}
\end{subfigure}
\begin{subfigure}[b]{0.45\textwidth}
\includegraphics[width=\textwidth]{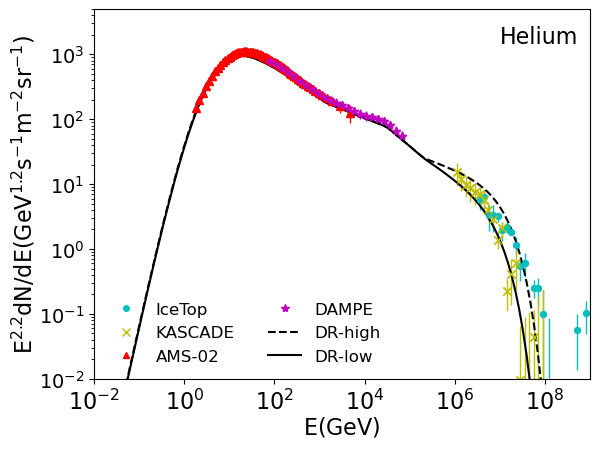}
\caption{helium}
\label{fig:elements:He}
\end{subfigure}
\captionsetup{justification=raggedright,width=0.45\textwidth}
\caption{Best-fitting spectra of protons (top), and helium nuclei (bottom), along with the observation data from IceTop (blue circles), KASCADE (yellow crosses), AMS-02 (red triangles), and DAMPE (purple stars). The solid line represents the DR-low model, while the dashed line represents the DR-high model.}
\label{fig:elements}
\end{figure}

One has to notice that the measurements of the energies of the knees disagree between IceTop and KASCADE, as shown in Figure~\ref{fig:elements}. As the KASCADE experiment uses the QGSJET-II-02 model while IceTop uses the Sibyll 2.1 model instead, the discrepancies are caused by the large systematic uncertainty of the hadronic model. In our study, we refer to the models derived from KASCADE and IceTop data as DR-low and DR-high respectively. 

\begin{figure}[!htp]
\includegraphics[width=0.45\textwidth]{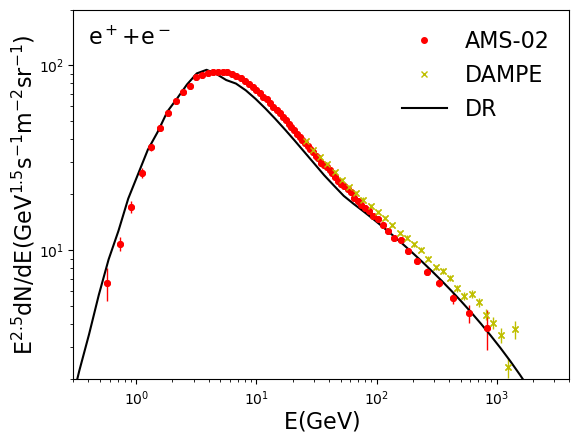}
\captionsetup{justification=raggedright,width=0.45\textwidth}
\caption{\label{fig:totep} The black solid line shows the fitted electrons plus positrons spectrum, and the measurements from AMS-02 (red dots) and DAMPE (yellow crosses) are also marked.}
\end{figure}

In this work, to estimate the diffuse gamma-ray and neutrino emission, the proton and electron plus positron spectra observed by Voyager, AMS-02, IceTop, and KASCADE are adopted to constrain the Galactic CR distribution as seen in Figure~\ref{fig:elements} and Figure ~\ref{fig:totep}.

\subsection{Gamma-ray sky}
The diffuse gamma-ray emission has been well measured by a few satellites below TeV energies, such as EGRET, and followed by Fermi-LAT~\cite{Hunter:1997qec, Fermi-LAT:2009ihh}. Recently, the Galactic plane has been observed up to 1 PeV, thanks to the Tibet AS$\gamma$ and LHAASO~\cite{TibetASgamma:2021tpz, LHAASO:2023gne} experiments. These discoveries show the evidence of hadronic origin of sub-PeV diffuse gamma rays, which are generated during the propagation of tens of PeV CRs.

The LHAASO experiment announced the source-subtracted galactic diffuse gamma-ray fluxes from the inner galactic plane (15$^{\circ}<$l$<125{^\circ}$, $|b|<5{^\circ}$) and outer plane (125${^\circ}<$l$<235{^\circ}$, $|b|<5{^\circ}$) for the first time lately. Where a simple power law is adopted to describe the spectra for both regions with similar spectral indices of $-2.99$, which is consistent with the CR spectral index of the knee region.

\begin{figure}
    \centering
    \includegraphics[width=0.45\textwidth]{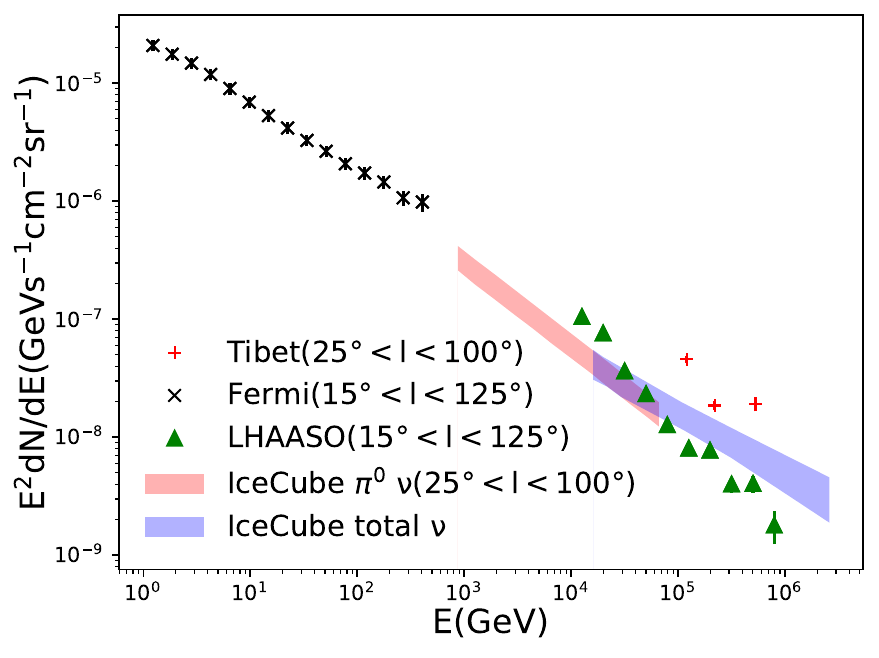}
    \captionsetup{justification=raggedright}
    \caption{The gamma-ray data from Fermi-LAT (black crosses) and LHAASO experiments (blue triangles) in the region of 15$^{\circ}<$l$<125{^\circ}$, $|b|<5{^\circ}$, gamma-ray data from Tibet AS$\gamma$ (red plus), IceCube total $\nu$ (blue shaded region) and their results with $\pi^0$ model in the Galactic plane (red shaded region) are shown.}
    \label{fig:gamma-neutrino}
\end{figure}

In Figure~\ref{fig:gamma-neutrino}, the data for the window of 15$^{\circ}<$l$<125{^\circ}$, $|b|<5{^\circ}$ from LHAASO and Fermi-LAT experiments \cite{zhangGalacticDiffuseGammaray2023}, and for 25$^{\circ}<$l$<100{^\circ}$, $|b|<5{^\circ}$ from Tibet AS$\gamma$ are presented. As can be seen, both the LHAASO and Tibet AS$\gamma$ data are in agreement with the Fermi-LAT data. However, the result from LHAASO is a few times lower than that from AS$\gamma$. This is due to the different analysis methods of these two experiments. Where LHAASO analyzes the data by masking sources included in the TeVCAT with a radius of five times the Gaussian extension widths. Therefore, this cut procedure may lose a large part data of the Galactic plane, where the diffuse CR and unresolved sources are located. 

\subsection{Neutrino sky}
Since the first observation of the astrophysical neutrino signal in the TeV - PeV energy range in 2012~\cite{IceCube:2013low}, IceCube has kept updating the neutrino sky for more than 10 years. The event distribution is consistent with being isotropic, and the origin of these neutrino signals is still unresolved. With larger statistics, IceCube has recently shown that there are more events at lower Galactic latitudes and a deficit of neutrino events at high Galactic latitudes~\cite{IceCube:2019lzm}. The IceCube Neutrino Observatory has provided 6-year all-sky total and 10-year Galactic plane data~\cite{IceCube:2019lzm, IceCube:diffuse}. In this recently updated data sample for Galactic neutrino search, they performed the analysis for the cascade events with lower energy thresholds. The neutrino emission from the Galactic plane is reported at the 4.5$\sigma$ level of significance~\cite{IceCube:diffuse} with a total of 59,592 events selected over the entire sky in the energy range of 500 GeV to several PeV. As shown in Figure~\ref{fig:gamma-neutrino}, the best-fitting Galactic plane neutrino flux is comparable with the gamma-ray flux.

The total neutrino observation as shown with blue shaded region in Figure~\ref{fig:gamma-neutrino} includes events from Galactic and extra-galactic diffuse backgrounds and astrophysical sources, whose spectrum follows a simple power-law distribution as can be found below, 
\begin{equation}\label{neutrino spectram}
    \Phi_{\nu}=\Phi_0 (\frac{E}{100 \mathrm{TeV}})^{-\gamma}.
\end{equation}
here the normalization factor $\Phi_0$ is 1.66$\times 10^{-18} \mathrm{GeV^{-1}cm^{-2}s^{-1}sr^{-1}}$, and the common spectral index $\gamma$ is 2.53.
The observed neutrino spectrum is softer than $E^{-2}$ which is comparable with the observed diffuse extragalactic gamma-ray background~\cite{bechtol2017evidence}.

\section{Models}\label{sec:models} 
Both the diffuse gamma-ray and neutrino flux are generated by CR particles when they propagate in the Galaxy.
The hadronic component of CRs can induce gamma-ray and neutrino emission through proton-proton interaction, as well as bremsstrahlung radiation. On the other hand, the leptonic component of CRs contributes to gamma-ray emissions through the inverse Compton (IC) effect.
To calculate these processes, along with the propagation effect of CRs, we utilize the well-established GALPROP code~\cite{Strong:1998pw} in this study.
In this section, we provide a detailed description of the CR propagation and emission model that are employed.

\subsection{Cosmic-ray propagation}

In the propagation model, CRs are assumed to undergo diffusion within the Galactic magnetic field, taking into account the possible effects such as reacceleration, energy loss, fragmentation, and decay.
The diffusion coefficient is parameterized as $D(R)=\beta^{\eta}D_0(R/4\mathrm{GV})^{\delta}$, where $D_0$ is the normalization factor at a reference rigidity of 4 GV, R is the particle's rigidity, $\beta$ is the velocity of the particles in natural units, $\delta$ is the slope of rigidity dependence, and $\eta$ is a phenomenological parameter introduced to fit the low energy secondary-to-primary ratios.
Besides the diffusion, the convection or reacceleration effect is also required by the observed $B/C$ data.

In some recent studies, with more secondary CR species like Li, Be, and B precisely measured by AMS-02, it was found that the reacceleration effect is favored~\cite{yuanSecondaryCosmicrayNucleus2020}.
In this work, we adopt a Diffusion plus Reacceleration (DR) model as a benchmark model. The model parameters that describe the propagation processes are adopted following the work of other groups~\cite{yuanSecondaryCosmicrayNucleus2020}, corresponding to the best-fit values obtained by fitting the Li, Be, B, C, and O measurements from AMS-02.
As listed in Table~\ref{tab:prop}, the half height of diffuse zone $z_h$ is 6.3 kpc, and the Alfven speed $v_A$ that describes the strength of reacceleration is $33.76~\mathrm{km ~s^{-1}}$.
\begin{table}[!htp]
\caption{Propagation parameters.}
\begin{tabular}{cccccc}
    \hline\hline
     $D_0$ & $\delta$ & $z_h$ & $v_A$ & $\eta$ \\
    & ($10^{28}$~cm$^2$~s$^{-1}$) &  (kpc) & (km~s$^{-1}$) & & \\
    \hline
     $7.69$ & $0.362$ & $6.3$  & $33.76$ & $-0.05$ \\
    \hline
\end{tabular}
\label{tab:prop}
\end{table}

\subsection{Cosmic-Ray injection}
It is widely accepted that SNRs are the most promising galactic high-energy CR sources, whose shock provides the ideal environment for first-order Fermi accelerations of relativistic particles. In the discovery of 12 Galactic PeV accelerators by LHAASO, eight of them are somehow linked to SNRs~\cite{LHAASO:2021gok}.
Therefore we make a simple assumption that CRs are injected into the Galaxy by the SNRs.
As it is not possible to gather information on all historical SNRs, for an estimation, we employ a continuous source distribution for SNRs as follows
\begin{equation}
    f(r,z)=\left(\frac{r}{r_{\odot}}\right)^{1.25}\exp\left[-\frac{3.56(r-r_{\odot})}{r_{\odot}}\right]\exp\left(-\frac{|z|}{z_s}\right),
\end{equation}
where $r_\odot=8.3\mathrm{kpc}$ is the distance of the sun, $z_s=0.2\mathrm{kpc}$ is a scale factor that indicates the thickness of the Galactic disk.

Given the many spectral structures revealed by recent direct detection experiments, the injection spectra of CRs may be quite complicated. A multiple-broken-power-law spectrum is employed to describe these features as seen below,

\begin{equation}
f(x) = \left\{
\begin{aligned}
      &R^{-\nu_0}e^{-\frac{R}{Rc}}, & R < R_1 \\
      &(\prod_{i=1}^{n}R_i^{\nu_i-\nu_{i-1}})R^{-\nu_n}e^{-\frac{R}{Rc}}, & R_n \leq R < R_{n+1} \\
      &(\prod_{i=1}^{4}R_i^{\nu_i-\nu_{i-1}})R^{-\nu_4}e^{-\frac{R}{Rc}}, & R_4 \leq R
    \end{aligned}
\right.
\label{eq:inj}
\end{equation}

where n in the second row is from 1 to 3. The corresponding observation could constrain the injection parameters in Equation~\ref{eq:inj} for different CR species. As there exist discrepancies between the IceTop and KASCADE measurements, we construct two models, differing in their injections, to indicate the upper and lower boundary of theoretical estimation, so-called DR-high and DR-low models respectively. 

Following Ref.~\cite{zhangGalacticDiffuseGammaray2023}, the spectral structures are assumed to be mainly due to the source injection as in Equation~\ref{eq:inj}, without taking into account the change of propagation parameters. 
The constraints for proton, helium, and electron plus positron were performed. These three kinds of CR particles play a major role in contributing to Galactic $\gamma$ and neutrino emission.
We develop our neutrino and gamma-ray sky map based on these best-fitting parameters, listed in Table~\ref{tab:inj_phe} and ~\ref{tab:inj_ep}. The comparisons between observation and model are shown in Figure~\ref{fig:elements} and ~\ref{fig:totep}. 

\begin{table}[!htb]
\small
\captionsetup{justification=raggedright,width=0.45\textwidth}
\caption{Source injection and solar modulation parameters as in Equation \ref{eq:inj} for proton and Helium nuclei.}
\begin{tabular}{c|cc|cc}
\hline\hline
   & \multicolumn{2}{c|}{Proton} & \multicolumn{2}{c}{Helium} \\

 & DR-high & DR-low & DR-high & DR-low \\ 
& (IceTop) & (KASCADE)&(IceTop)&(KASCADE)\\ \hline
$\nu_0$ &$2.06$&$2.06$ &$1.46$&$1.46$ \\
$\nu_1$ &$2.43$&$2.43$ &$2.36$&$2.36$ \\
$\nu_2$ &$2.22$&$2.22$ &$2.12$&$2.12$ \\
$\nu_3$ &$2.52$&$2.52$ &$2.42$&$2.42$ \\
$\nu_4$ &$2.18$&$2.32$ &$2.08$&$2.28$ \\
$R_1/{\rm GV}$ &$13.9$&$13.9$ &$1.99$&$1.99$ \\
$R_2/{\rm TV}$ &$0.50$&$0.50$ &$0.65$&$0.65$ \\
$R_3/{\rm TV}$ &$15.0$&$15.0$ &$15.0$&$15.0$ \\
$R_4/{\rm TV}$ &$100.0$&$100.0$ &$100.0$&$100.0$ \\
$R_c/{\rm PV}$ &$12.0$&$4.0$ &$6.0$&$4.0$ \\
$\Phi/{\rm GV}$&$0.700$&$0.700$ &$0.700$&$0.700$ \\
\hline
\end{tabular}
\label{tab:inj_phe}
\end{table}

\begin{table*}
\small
\caption{Source injection and solar modulation parameters of electron plus positron.}
\begin{tabular}{cccccccccc}
\hline
 $\nu_0^-$ & $\nu_1^-$ & $\nu_2^-$ & $\nu_3^-$ & $R_1^-$/GV & $R_2^-$/GV & $R_3^-$/GV &$R_c^-$/TV & $\Phi^-$/GV \\
 $2.33$ & $0.01$ & $2.88$ & $2.45$ & $0.950$ & $4.19$ & $55.7$ & $6.27$ & 1.1\\
\hline
 $c_{e^+}$ & $\nu_1^+$ & $\nu_2^+$ & $R_1^+$/GV & $R_c^+$/TV & $\Phi^+$/GV \\
 $1.00$ & $3.04$ & $2.08$ & $31.2$ & $3.42$ & 1.1 \\
\hline
\end{tabular}\\
\label{tab:inj_ep}
\end{table*}

As seen in Table \ref{tab:inj_phe}, for both DR-high and -low model, most of the best-fitting parameters $\nu_i$ and $R_i$ are identical to each other, except $R_c$ and $\nu_4$.
Here $R_c$ represents the characteristic cutoff rigidity of the exponential cutoff spectral, describing the knee energy of those particles. Apparently, the $R_c$ of DR-high is higher than that of DR-low, because of the different knee energies from IceTop and KASCADE. On the other hand, the $\nu_4$ of DR-high is smaller than that of DR-low, which is due to a harder spectrum from the IceTop measurement.

\subsection{Gamma-Ray expectation} \label{subsec:sky_map_expectation}

With the propagation and injection of CRs fixed, we analyze the gamma-ray sky map. We apply the GALPROP code to calculate the diffuse emissions from a few processes, including natural pion decay, bremsstrahlung, and inverse Compton scattering (ICS).
The AAfrag package~\cite{Kachelriess:2019ifk} is adopted to estimate the secondary gamma-ray and neutrino production from inelastic hadronic interactions.

\begin{figure*}
\begin{subfigure}[b]{0.45\textwidth}
\includegraphics[width=\textwidth]{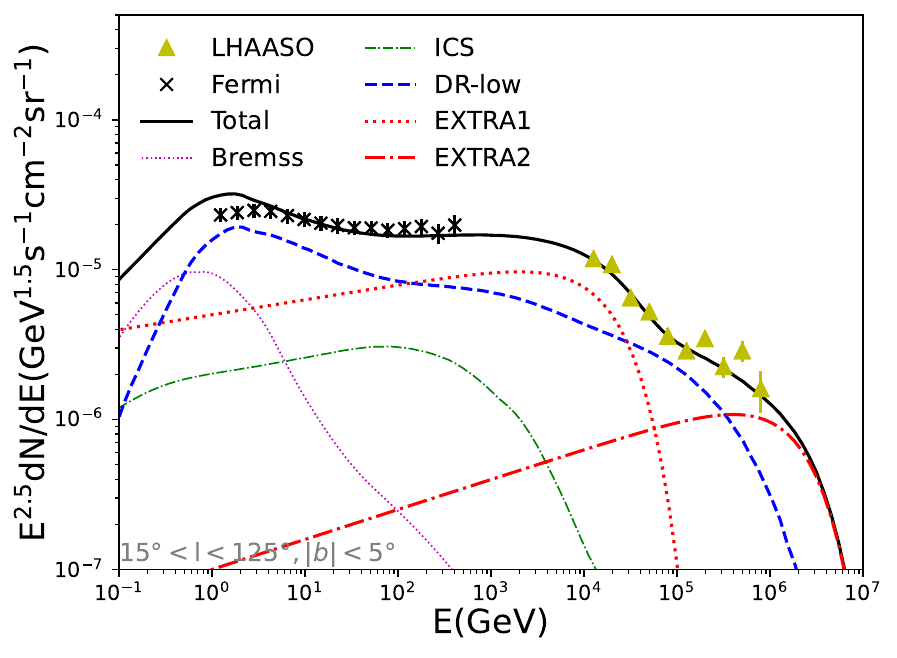}
\caption{Inner region $\&$ DR-low}
\label{fig:gamma:inner_low}
\end{subfigure}
\begin{subfigure}[b]{0.45\textwidth}
\includegraphics[width=\textwidth]{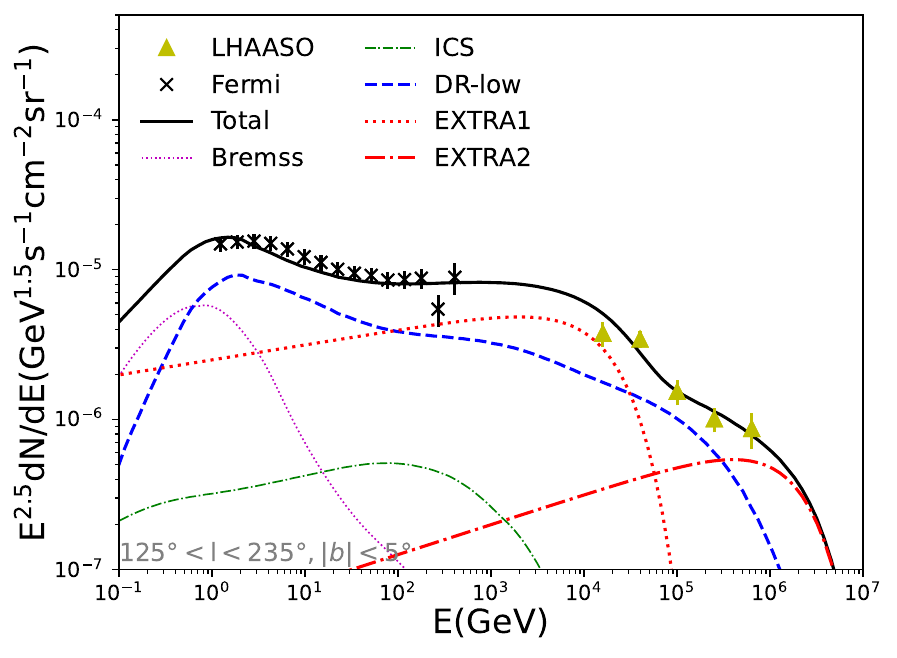}
\caption{Outer region $\&$ DR-low}
\label{fig:gamma:outer_low}
\end{subfigure}
\begin{subfigure}[b]{0.45\textwidth}
\includegraphics[width=\textwidth]{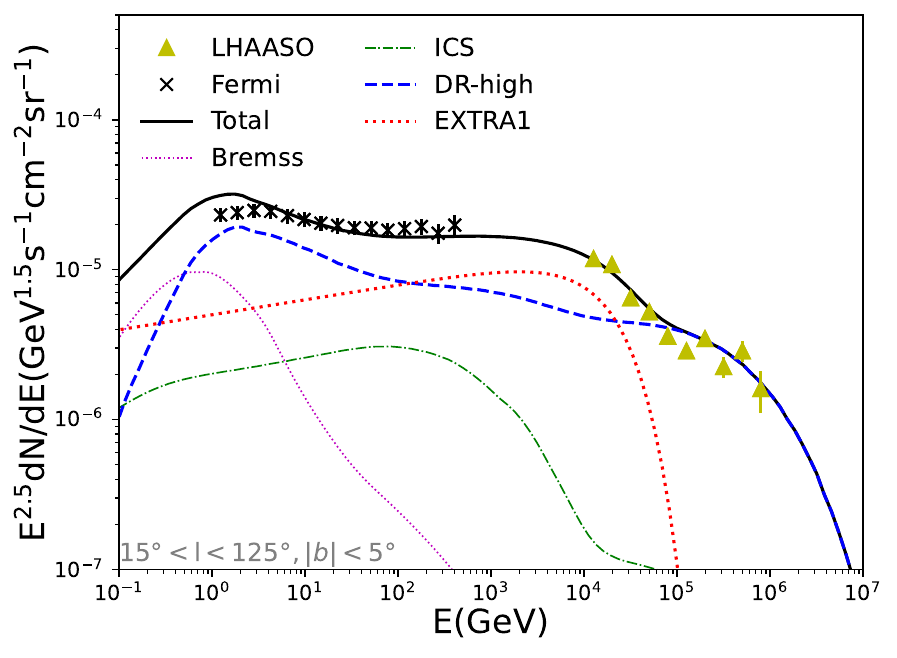}
\caption{Inner region $\&$ DR-high}
\label{fig:gamma:inner_high}
\end{subfigure}
\begin{subfigure}[b]{0.45\textwidth}
\includegraphics[width=\textwidth]{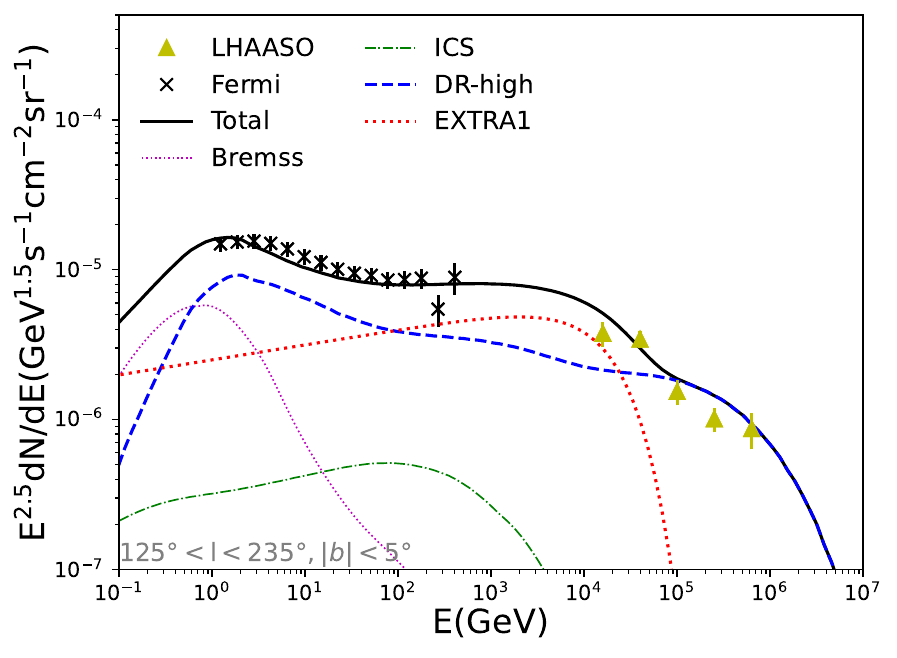}
\caption{Outer region $\&$ DR-high}
\label{fig:gamma:outer_high}
\end{subfigure}
\captionsetup{justification=raggedright,width=0.94\textwidth}
\caption{\label{fig:gamma} The diffuse gamma-ray emission calculated from the DR model. The physical radiation of ICS (green dot-dashed line), bremsstrahlung (pink dotted line), and pion decay (blue dashed line) are shown. Two extra source components, EXTRA1 (red dotted line) and EXTRA2 (red dot-dashed line) with ECPL spectra are presented. Panel (a) and (b) are the spectra obtained from the DR-low model and panel (c) and (d) for the DR-high model. The (a) and (c) panels show the results for the inner Galaxy region of 15$^{\circ}<$l$<125{^\circ}$, $|b|<5{^\circ}$, while the (b) and (d) panels display the results for the outer Galaxy region of 125${^\circ}<$l$<235{^\circ}$, $|b|<5{^\circ}$.}
\end{figure*}

We show the diffuse gamma-ray spectra measured by the LHAASO and Fermi-LAT experiments, along with our model predictions for both the inner region (Figure~\ref{fig:gamma:inner_low} and \ref{fig:gamma:inner_high}) and outer region (Figure~\ref{fig:gamma:outer_low} and \ref{fig:gamma:outer_high}).
To ensure a self-consistent comparison, we apply the same masks as in the LHAASO analysis~\cite{LHAASO:2023gne} for all calculated results and data.

Compared with the gamma-ray data from Fermi-LAT and LHAASO, 
the predicted flux with the DR-high-only model is consistent with the data both at energies less than a few GeV and above 60 TeV. However, between a few GeV and 60 TeV, DR-high-only model can not explain the LHAASO data, as can be seen in Figure ~\ref{fig:gamma:inner_high} and ~\ref{fig:gamma:outer_high}.

This excess below 60 TeV was initially identified through the analysis of GeV Fermi-LAT observations~\cite{ackermann2012fermi}.
To account for this excess, some studies have proposed a spatially dependent diffusion model~\cite{guoUnderstandingSpectralHardenings2018}.
However, this modification of the propagation model is insufficient to explain the data obtained by LHAASO.

In this work, we attribute this TeV excess to unresolved sources along the Galactic plane, which are expected to be numerous and faint within the field of view of LHAASO and Fermi-LAT. Various physical interpretations have been previously discussed in the literature~\cite{kantzasPossibleContributionXray2023, lindenPulsarTeVHalos2018, vecchiottiUnresolvedSourcesNaturally2022}.
Among these interpretations, the pulsar TeV halo and pulsar wind nebulae (PWNe) have emerged as potential candidates~\cite{lindenPulsarTeVHalos2018,vecchiottiUnresolvedSourcesNaturally2022}.
Therefore to fit the bump $\sim\mathcal{O}(1)\mathrm{TeV}$ in the LHAASO spectrum, we employ an ECPL component (named EXTRA1) with an index of 2.4 and a cutoff of 20 TeV to describe these unresolved sources that follow the spatial distribution of pulsars.

The cutoff energy of the introduced EXTRA1 in this work is lower than that of the extra component in Reference~\cite{zhangGalacticDiffuseGammaray2023} (30 TeV), as we have introduced the EXTRA2 to account for the high-energy data.

However, this component is insufficient for the DR-low case, where an additional component is required at PeV energy.
Therefore, another ECPL component (EXTRA2) with an index of 2.3 and a cutoff at 2 PeV is introduced for the DR-low case. These EXTRA1 and EXTRA2 components have close spectral indices, around the average spectral index of sources in the H.E.S.S. Galactic Plane Survey, but different cutoff energies. The cutoff energies of gamma rays for different scenarios indicate different maximum energies of CR particles. For instance, the leptonic (hadronic) origin, the 20 TeV gamma-ray cutoff energy corresponds to 700 TeV (100 TeV) electron/positron (proton) cutoff energy. This suggests that EXTRA1 and EXTRA2 likely represent at least two distinct types of unresolved sources in the Galaxy. 

Even though recent studies indicate that this excess, so as our EXTRA1 components, prefers to be leptonic origin, as strongly constrained by the hardening of the local cosmic-ray proton spectrum observed by AMS. No source class has been uncovered. 

For EXTRA2 sources contributing at higher energies, there is no constrain from current cosmic-ray observation. If they are of leptonic origin, such as PWNe, the Klein-Nishina regime is dominant. Therefore a very high acceleration rate is required and should also be higher than the electron radiative losses. This is quite stringent. If EXTRA2 sources are TeV halos, some studies argued a slower diffusion of the electrons in the interstellar medium is needed \cite{PhysRevLett.126.241103}, which is still not understood. Therefore, the hadronic model cannot be excluded. To confirm and deep explore the source mechanisms of both EXTRA1 and EXTRA2, neutrino signals as the smoking gun would provide the direct evidence for this mystery.

\section{Results}\label{resutls}

\subsection{Galactic diffuse gamma-ray emission}

\begin{figure}[!htb]
\includegraphics[width=0.45\textwidth]{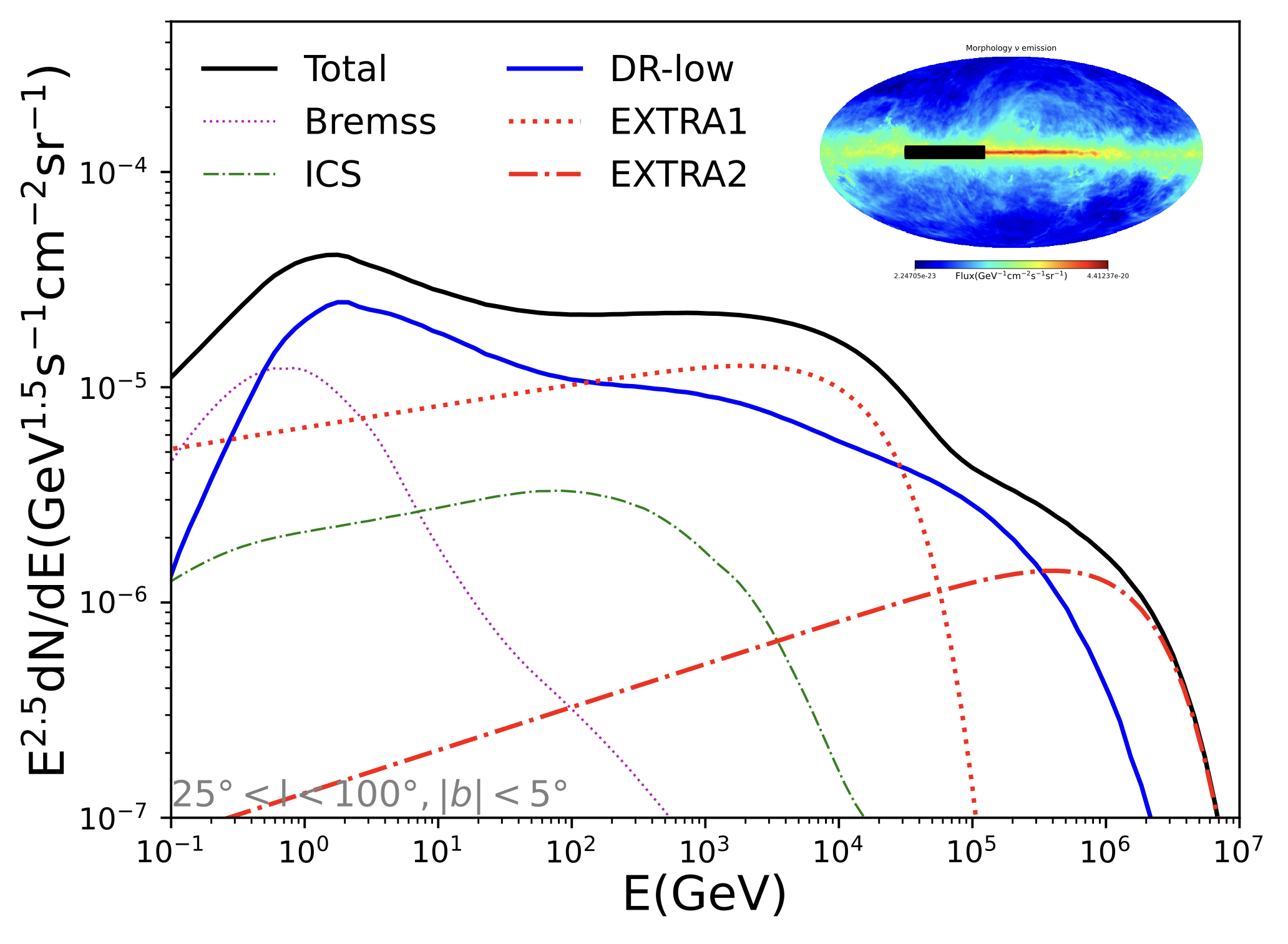}
\captionsetup{justification=raggedright,width=0.45\textwidth}
\caption{\label{fig:gamma_25_100} The diffuse gamma-ray emission calculated from the DR-low model. The physical radiation of ICS (green
dot-dashed line), bremsstrahlung (pink dotted line), and pion decay (blue dashed line) are shown. Two extra source
components, EXTRA1 (red dotted line) and EXTRA2 (red dot-dashed line) with ECPL spectra are presented. This figure shows the result for the inner Galaxy region of 25$^{\circ}<$l$<100{^\circ}$, $|b|<5{^\circ}$}.
\end{figure}

Based on the constructed model, we generate a diffuse gamma-ray emission map which can be used as a template for future studies.
This map consists of four components: ICS, bremsstrahlung, natural pion decay, and extra source contributions.
Except for bremsstrahlung and neutral pion decay, the spatial distributions of all these components are different from each other. 
In Figure~\ref{fig:gamma_25_100}, we show the gamma-ray energy spectrum for the region of 25$^{\circ}<$l$<100{^\circ}$, $|b|<5{^\circ}$ without masking as an example. In general, this spectrum is higher than that of the region 15$^{\circ}<$l$<125{^\circ}$, which might be due to the masking effect from the LHAASO analysis.
For any other region of interest, the predicted gamma-ray emission can be selected in the same manner, to serve as a background template for point source analysis.

\subsection{Galactic diffuse neutrino}

\begin{figure}[!htb]
\includegraphics[width=0.45\textwidth]{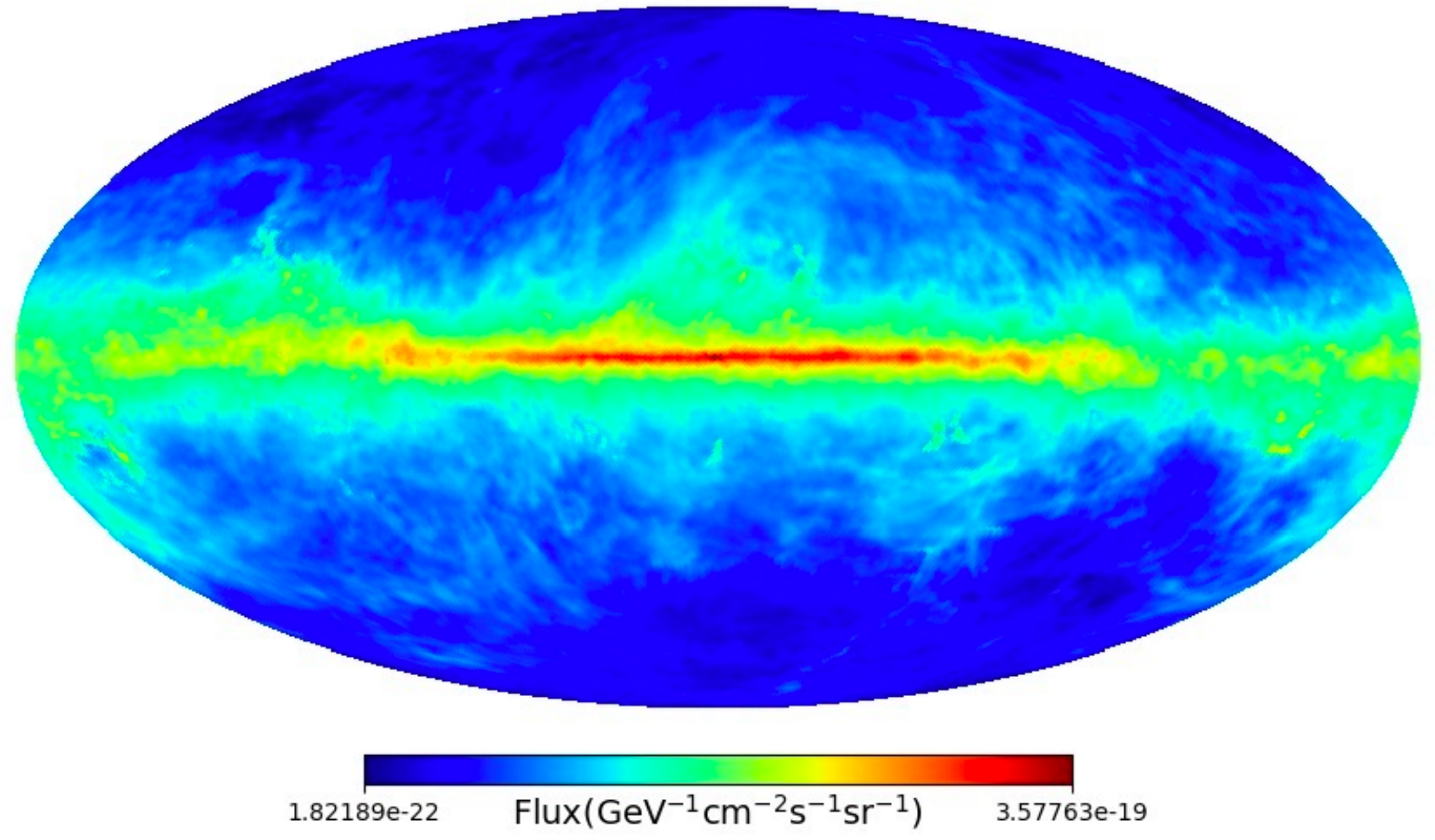}
\captionsetup{justification=raggedright,width=0.45\textwidth}
\caption{\label{fig:nu-map} Calculated galactic diffuse neutrino map with energies from 100 TeV to 10 PeV. The morphology follows the gas distribution in our Galaxy.}
\end{figure}

\begin{figure}[]
\centering
\begin{subfigure}[b]{0.45\textwidth}
\includegraphics[width=\textwidth]{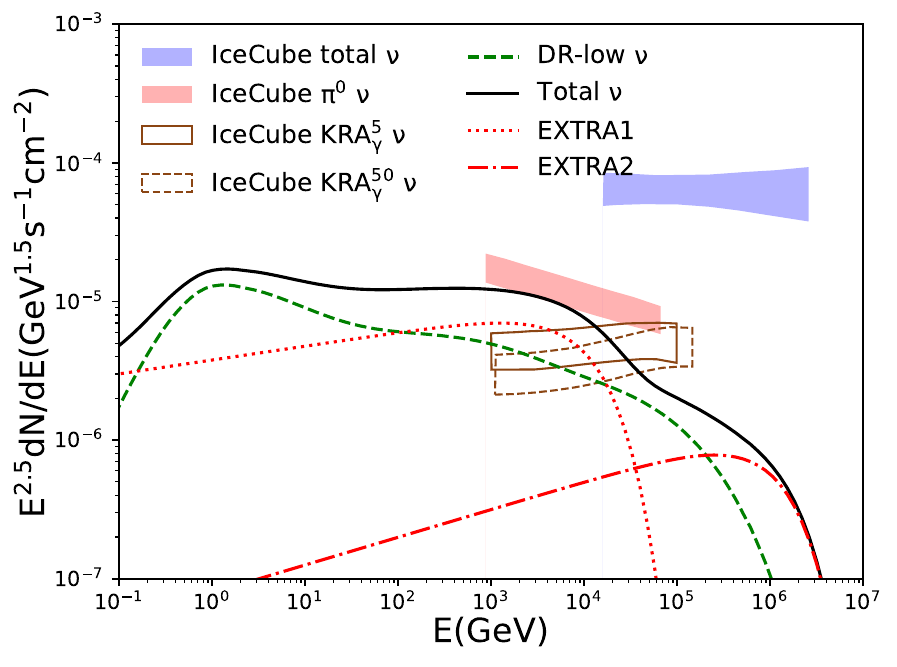}
\caption{all-sky}
\label{fig:nu-spectrum:allsky}
\end{subfigure}
\begin{subfigure}[b]{0.45\textwidth}
\includegraphics[width=\textwidth]{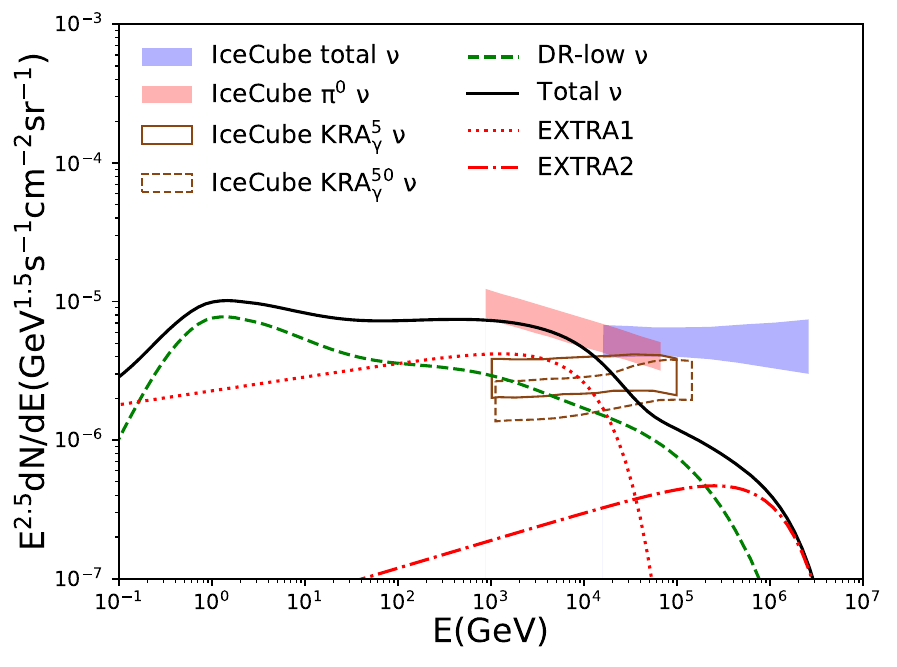}
\caption{25$^{\circ}<$l$<100{^\circ}$, $|b|<5{^\circ}$}
\label{fig:nu-spectrum:25_100}
\end{subfigure}
\captionsetup{justification=raggedright,width=0.45\textwidth}
\caption{\label{fig:nu-spectrum} The predicted neutrino flux per flavor from the DR-low model compared with the IceCube total data (blue shaded region), their $\pi^0$ model (red shaded region), $KRA_\gamma^5$ model (region with brown solid edge) and $KRA_\gamma^{50}$ model (region with brown dashed edge). Other components including EXTRA1 (red dotted line), EXTRA2 (red dot-dashed line), neutrino flux with DR-low model (green dashed line), and total $\nu$ flux (black solid line) are shown. Panel (a) is for the all-sky region, and panel (b) is in the region of 25$^{\circ}<$l$<100{^\circ}$, $|b|<5{^\circ}$.}
\end{figure}

We show the neutrino sky map from 100 TeV to 10 PeV resulting from Section~\ref{sec:models} in Figure~\ref{fig:nu-map}.
As one can see in Figure~\ref{fig:nu-spectrum:allsky} and Figure~\ref{fig:nu-spectrum:25_100}, our prediction for Galactic diffuse neutrino emission for both all-sky and Galactic plane with DR-low model are in agreement with IceCube best-fitting flux normalizations from the data~\cite{IceCube:diffuse}. 
However, for the $\pi^0$ template of IceCube, an extra source contribution with a hadronic origin is needed. This appears to contradict the fact that these sources only contribute to gamma-ray emissions and not cosmic rays. 

For comparison, IceCube's total neutrino is also shown here.
Our calculated Galactic diffuse neutrino flux shows that the contribution of Galactic neutrinos to the total neutrino observation is around $9\%$ at 20 TeV, as seen in Figure~\ref{fig:nu-spectrum:allsky}.

In Figure~\ref{fig:nu-spectrum:25_100}, we present a comparison of the surface brightness of one flavor neutrino between the Galactic contribution in the disk region ($|b|<5^\circ$, $25^\circ<l<100^\circ$) and the total contribution averaged over the all-sky region. 
This shows the distinctiveness of the neutrino Galactic disk compared to the isotropic neutrino background. 
The neutrino flux of the Galactic disk is prominent in the energy range from 10 TeV to 100 TeV and decreases significantly at higher energies. This is constrained by the gamma-ray and cosmic-ray measurements. Our results are in agreement with other groups' study~\cite{IceCube:diffuse, Kovalev:2022izi}. The Milky Way is a source of high-energy neutrinos consistent with the gamma-ray observation, as seen in Figure~\ref{fig:nu-spectrum:allsky} and ~\ref{fig:gamma_25_100}. 

In the case of DR-high-only model, as seen in Figure \ref{fig:dr-high-extra1-neutrino}, calculated neutrino flux is consistent with two best-fitting results with KRA$_\gamma$ model. At a few PeV, the Glashow resonance is shown in the spectrum \cite{Glashow:1960zz}. However, to explain the results for the $\pi^0$ model, EXTRA1 would be necessary. 

\section{Discussion}\label{discussion}

In this work, based on the most recent PeV Galactic diffuse gamma-ray observation from LHAASO, with two sets of CR data from IceTop and KASCADE, we construct our DR-high and DR-low models separately. For both models, we find it is hard to explain the LHAASO Galactic plane search with conventional CR propagation. After adding extra source contribution, the diffuse gamma-ray emission can be well explained both by DR-high with EXTRA1 (Model 1) and DR-low with both EXTRA1 and EXTRA2 (Model 2).

For Model 1, one extra source spectrum EXTRA1 is introduced, with a spectral index of 2.4 and 20 TeV cutoff. For Model 2, two extra source contributions are introduced, where one has an index of 2.4 and 20 TeV cutoff energy and another with an index of 2.3 and 2 PeV cutoff energy. That means there could be two populations of sources in our Galaxy with faint gamma-ray emission which is lower than the sensitivity of our current instruments, so they have not been identified. They follow similar CR accelerated mechanisms with close spectral index, but various maximum CR energy. 

Based on the obtained model, we simulated the Galactic diffuse neutrino flux, obtaining the sky map as shown in Figure~\ref{fig:nu-map}. For example with Model 2, we estimate the Galactic contribution of the astrophysical flux is around 9 $\%$ at 20 TeV. It is uncertain if these Galactic neutrinos are from the CR propagation or point sources because of insufficient statistical power. Therefore, we believe the future Imaging Air Cherenkov Telescope ~\cite{Marinos:2022tdj} and upgraded neutrino observatory will resolve the point sources and precisely provide the diffuse map and reveal the origin and propagation of cosmic rays. 

The best-fitting Galactic neutrino flux from IceCube is model dependent. Where the $\pi^0$ model is constrained by the Fermi MeV to GeV gamma-ray emission and is extrapolated to TeV, where the same spatial emission profile is assumed. While KRA$_\gamma$ models take into account the spatial distribution of spectra, and cutoff energies of 5 and 50 PeV respectively. Therefore, the $\pi^0$ model provides an even event distribution along the Galactic plane, and KRA$_\gamma$ models give a higher neutrino flux at the Galactic center region. So that for the interested region of 25$^{\circ}<$l$<100{^\circ}$, $|b|<5{^\circ}$, the $\pi^0$ model gives higher flux than that from KRA$_\gamma$ models. On the other hand, the cosmic-ray diffuse modeling with Galprop for the DR-low and DR-high models in this work, is not consistent with the KRA models from the Dragon analysis. The discrepancy between all these models is due to the low statistics and the uncertainty of the current templates. Further accurate measurements and studies are quite essential. We summarize the differences in Table~\ref{tab:extra-dis}.

\begin{figure}[!htb]
\includegraphics[width=0.45\textwidth]{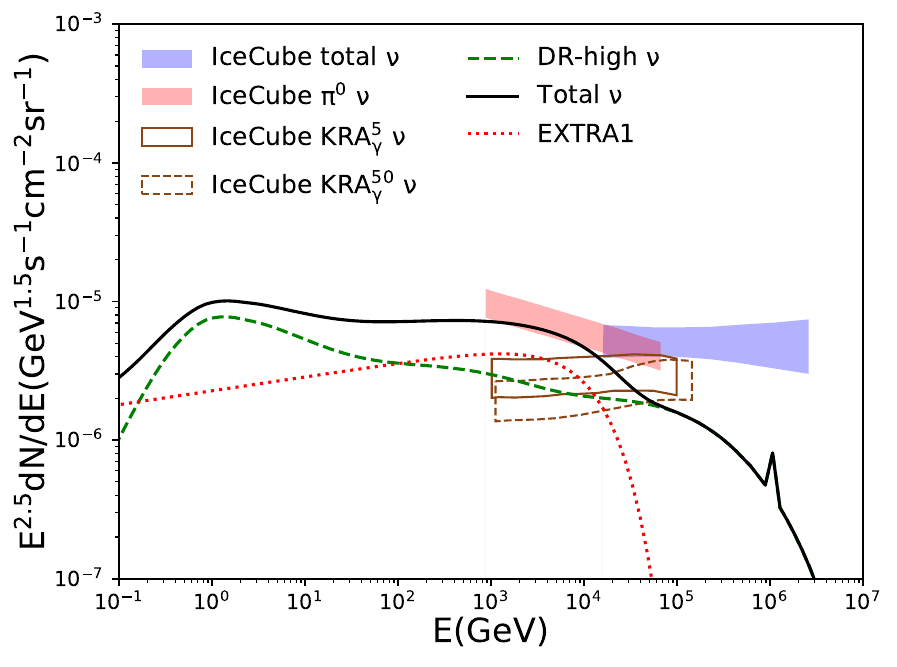}
\captionsetup{justification=raggedright,width=0.45\textwidth}
\caption{\label{fig:dr-high-extra1-neutrino} The predicted neutrino flux per flavor for the DR-high model compared with the IceCube total data (blue shaded region), their $\pi^0$ model (red shaded region), $KRA_\gamma^5$ model (region with brown solid edge) and $KRA_\gamma^{50}$ model (region with brown dashed edge). Neutrino flux from EXTRA1 (red dotted line), DR-high model (green dashed line), and total $\nu$ flux (black solid line) are shown for the region of 25$^{\circ}<$l$<100{^\circ}$, $|b|<5{^\circ}$.}
\end{figure}

With only the gamma-ray and cosmic-ray observation, EXTRA1 sources prefer to be the leptonic origin, which has been discussed by a few groups. However, in the case of IceCube best-fitting flux for the $\pi^0$ model which is the only one consistent with the recent observations of 100 TeV gamma rays by the Tibet AS$\gamma$ \cite{TibetASgamma:2021tpz}, a population of EXTRA1 sources with a hadronic scenario would be necessary no matter of DR-high or DR-low model, as seen in Figure \ref{fig:nu-spectrum} and \ref{fig:dr-high-extra1-neutrino}. It would require this kind of source to inject fewer high-energy protons. So the identification of neutrinos can reveal the origin of CRs, modify the CR propagation and distribution models drastically and explore the history of our Galaxy. If EXTRA1 sources are the leptonic origin, where no neutrino is produced. A tension exists between the predicted diffuse Galactic neutrino flux and the IceCube results for the $\pi^0$ model. 

For the case of IceCube best-fitting fluxes for the KRA$_\gamma$ models, it provides the lower limit for the neutrino emission from the Galactic plane. No other extra hadronic scenario sources are needed. In other words, the EXTRA1 sources would be of the leptonic origin. Which needs no extra proton injection and releases the tension between the data and models. 

No matter which results for the different model templates obtained by IceCube, both leptonic and hadronic origins of EXTRA2 source are allowed by data. High-energy neutrino emission is a unique diagnostic of hadronic content. With the future PeV neutrino detection with improved sensitivity, the EXTRA2 sources could be identified.
If EXTRA2 of Model 2 is the hadronic origin, the Galactic neutrino will contribute around $1\%$ to the total IceCube neutrino at PeV. Otherwise, the contribution is $\sim$ $0.4\%$.

\begin{table}[!htp]
\captionsetup{justification=raggedright,width=0.5\textwidth}
\caption{The scenario of extra source meets the measurements of cosmic rays, gamma rays, and neutrinos.}
\begin{tabular}{cccc}
    \hline\hline
    Origin & $p$ & $\nu$ & $\gamma$\\
    \hline
    Leptonic & $\checkmark$ & KRA$_\gamma$ & $\checkmark$ \\
    Hadronic & hadron-less & $\pi^0$   & $\checkmark$ \\
    \hline
\end{tabular}
\label{tab:extra-dis}
\end{table}
\section{Summary}\label{summary}
In summary, thanks to the recent observations from LHAASO \cite{LHAASO:2023gne} and IceCube \cite{IceCube:diffuse}, the Galactic diffuse sky has become richer, especially in the high-energy regime. The LHAASO measurements show a bump in the gamma-ray spectrum, where contributions from extra unresolved sources are needed. The IceCube Collaboration confirms the high-energy neutrinos from the Galactic plane. Our calculated flux with models obtained from gamma-ray observation is consistent with the neutrino data. However for the best-fitting results for the $\pi^0$ model from IceCube data, the EXTRA1 sources with hadronic scenario is a must to fill the gap between calculated flux and data. Even though it would be disfavored by CR measurements.

The joint analysis of cosmic rays, gamma rays, and neutrinos has shown strong power in understanding the high-energy sky. 
For example, the diffuse gamma-ray detection by LHAASO can probe the CR density in our Galaxy and solve the problem of the disagreement between IceTop and KASCADE. Secondly, the neutrino detection can reveal the hidden sources which are not transparent for gamma-ray emission.

The current results from all these three messengers are in agreement with each other. More evidence shows the existence of PeVatrons in our Galaxy. The next step forward should be identifying the mysterious astronomical origin of high-energy cosmic rays with upgraded neutrino and gamma-ray detectors. 

\section{Acknowledgements}
We thank the referee for the useful and helpful comments and suggestions. This work is supported by the National Natural Science Foundation of China (NSFC) grants 12005313, 12205388, and 12261141691.




\nocite{*}

\bibliography{apssamp}

\end{document}